\documentclass[twocolumn,showpacs]{revtex4}
\usepackage{amsmath}
\usepackage{amssymb}
\usepackage{bm}
\usepackage{epsfig}
\usepackage{graphicx}
\usepackage{color}

\def\dfrac{\displaystyle\frac}

\begin{document}

\title{Fine structure of two-electron states in single and double quantum dots}

\author{M.M. Glazov}

\affiliation{Ioffe Physical-Technical Institute RAS, 194021 St.-Petersburg, Russia}


\begin{abstract}
Energy spectrum fine structure of triplet two-electron states is investigated theoretically.
Spin-orbit interaction induced terms in the effective Hamiltonian of electron-electron interaction are derived for zinc-blende lattice semiconductor nanostructures: quantum wells and quantum dots. The effects of bulk and structural inversion asymmetry are taken into account. Simple analytical expressions describing the splittings of the two-electron states localized in a single quantum dot and in a lateral double quantum dot are derived. The spin degeneracy of triplet states is shown to be completely lifted by the spin-orbit interaction. An interplay of the conduction band spin splitting and the spin-orbit terms in electron-electron interaction is discussed. The photoluminescence spectra of hot trions and of doubly charged excitons are calculated and are shown to reveal the fine structure of two-electron states.
\end{abstract}

\date{\today}

\maketitle

\section{Introduction}

Semiconductor quantum dots (QDs) also known as artificial atoms demonstrate unique spin-dependent properties and have a strong potential for the spintronics~\cite{ssc:optor}. The fine structure of optical emission spectra of QDs provides detailed information on the charge carriers energy spectrum and their interactions~\cite{ivchenko05a}.

The fine structure of neutral QDs energy spectrum and, hence, their luminescence spectra is determined by the exchange interaction between an electron and a hole forming a zero-dimensional exciton~\cite{ivchenko05a}. The situation  is different in the case of doped QDs: for instance, in $n$-type singly charged QDs the emission is dominated by $X^-$ trion consisting of two electrons and a hole, and in doubly charged QDs the emission involves the transition from the state with three electrons and one hole ($X^{2-}$) into the state with two electrons~\cite{bracker08,kusraev08,cortez02,ediger:036808}. In these occasions, the fine structure of optical spectra is controlled also by the exchange interaction between electrons.

It is well known that two-electron states are split into spin singlet and triplet states by the exchange Coulomb interaction. In the absence of the spin-orbit interaction, the triplet states are degenerate. It was demonstrated recently that the spin-orbit contribution to electron-electron interaction (\emph{spin-orbit exchange interaction}) splits two-electron triplet  state into three sublevels, one of which corresponds to zero projection of the total spin on the growth axis and two others correspond to linear combinations of the states with total spin projections being $\pm 1$, similar to those of localized excitons~\cite{glazov2009}. 

Here we extend the theory developed in Ref.~\cite{glazov2009} to allow for the full microscopic symmetry of real QDs made of zinc-blende lattice semiconductors. The effects of the bulk inversion asymmetry caused by the absence of an inversion center in the point symmetry group of a zinc-blende lattice and of the structural inversion asymmetry on the fine structure of two-electron triplet states are analyzed. The developed theory is applied to calculate the two-electron states and photoluminescence spectra of a hot trion and $X^{2-}$ complex in single and double QDs.

\section{Model}

In what follows we consider quantum disks (quantum well QDs) and lateral double QDs grown of zinc-blende lattice direct-gap materials. It is convenient to (i) obtain the spin-orbit contributions to an effective Hamiltonian of electron-electron interaction in a quantum well and (ii) calculate the fine structure of the two localized electron states using the derived Hamiltonian.

\begin{figure}[hptb]
 \includegraphics[width=\linewidth]{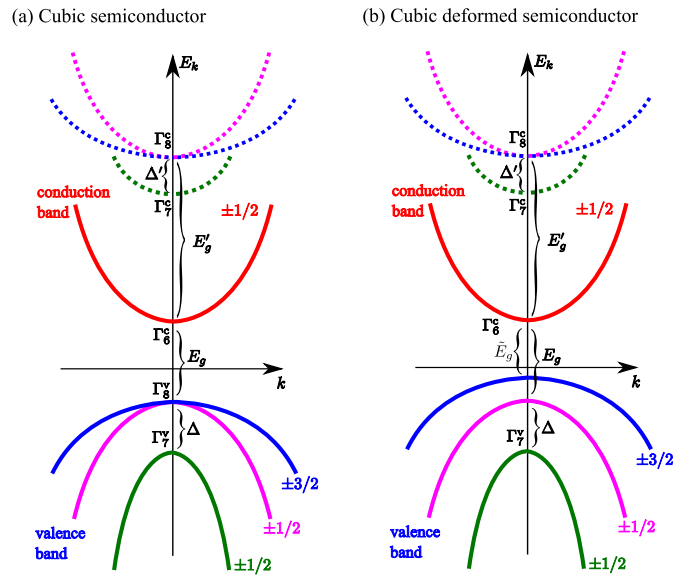}
\caption{(a) Schematic illustration of the band structure of a direct band zinc-blende lattice based semiconductor. (b) Schematic illustration of the band structure of a uniaxially deformed zinc-blende lattice based semiconductor. Splittings are shown not to scale.}
\label{fig:kane14}
\end{figure}

The schematic band structure of the direct band zinc-blende lattice semiconductor is depicted in Fig.~\ref{fig:kane14}(a). The spin-orbit interaction for the conduction band ($\Gamma_6^{\mathrm c}$) electrons is largely determined by the $\bm k \cdot \bm p$ admixture of the valence band states ($\Gamma_8^{\mathrm v}$, $\Gamma_7^{\mathrm v}$). The effects due to the lack of an inversion center in the point symmetry group of the bulk material can be taken into account by including into consideration the remote conduction bands $\Gamma_{8}^{\mathrm c}$, $\Gamma_{7}^{\mathrm c}$.  

Under the assumption that the gap $E_g'$ between the $\Gamma_6^{\mathrm c}$ and $\Gamma_{8}^{\mathrm c}$ exceeds by far the band gap $E_g$ (between $\Gamma_6^{\mathrm c}$ and $\Gamma_8^{\mathrm v}$) and that the spin-orbit splitting $\Delta'$ of the $\Gamma_{8}^{\mathrm c}$, $\Gamma_{7}^{\mathrm c}$ bands is much smaller as compared with $E_g'$, the band structure can be treated within the framework of the extended 8-band Kane model by including quadratic in the wave vector terms into the off-diagonal matrix elements characterizing the $\bm k \cdot \bm p$ interaction of the valence band states with the conduction band states~\cite{ivchenkopikus}. 

Free electron wave function in a quantum well grown along $z\parallel [001]$ axis can be represented as~(c.f. \cite{glazov2009})
\begin{equation}
 \label{kane}
\Psi_{s,\bm k}(\bm \rho,z) = e^{\mathrm i \bm k\bm \rho} [ \mathsf S_{\bm r} + \mathrm i {\mathbf R}_{\bm r} \cdot (A \hat{\bm \kappa}_{\bm K} - \mathrm i B \hat{\bm \sigma} \times \hat{\bm \kappa}_{\bm K})] \varphi(z) \, |  \chi_s\rangle.
\end{equation}
Here $\bm r = (\bm \rho,z)$ is the electron position vector,  $\varphi(z)$ is the smooth envelope describing its size quantization along $z$ axis, $\hat{\bm \sigma}$ is the electron spin operator, $\mathsf S_{\bm r}$ and $\mathbf R_{\bm r} = (\mathsf X_{\bm r}^{\mathrm v}, \mathsf Y_{\bm r}^{\mathrm v}, \mathsf Z_{\bm r}^{\mathrm v})$ are $s$-type and $p$-type Bloch functions of the conduction and valence bands, respectively, taken at the $\Gamma$ point, $|\chi_s\rangle$ is a spinor, $A=P(3E_g+2\Delta)/[3E_g(E_g+\Delta)]$ and $B=-P\Delta/[3E_g(E_g+\Delta)]$, $\Delta$ is the gap $\Gamma_8^{\mathrm v}$---$\Gamma_7^{\mathrm v}$, $P$ is Kane parameter. Equation~\eqref{kane} is valid provided that the electron energy referred to the conduction band bottom is much smaller as compared to $E_g$ and $\Delta$.

Vector $\hat{\bm \kappa}_{\bm K}$ describing the admixture of the valence bands has the following Cartesian components in the cubic axes $x\parallel [100]$, $y\parallel [010]$
\begin{equation}
 \label{kappa:vec}
\hat{\bm \kappa}_{{\bm K},i} = \hat{ K}_i  - \mathrm i \beta \hat{ K}_{i+1} \hat{ K}_{i+2},
\end{equation}
where $\hat{\bm K} = (\bm k, -\mathrm i \partial/\partial z)$ is the wave vector of the electron, a cyclic rule is applied for the subscripts ($i+3=i$) and constant $\beta$ can be related to the bulk Dresselhaus parameter $\gamma_{\rm c}$ describing cubic in $k$ conduction band spin splitting as $\beta= \gamma_{\rm c}/2BP$~\cite{ivchenkopikus,Knap}. 

The matrix elements of the effective electron-electron interaction Hamiltonian taken between the states $(\bm k s, \bm k's')$ and $(\bm p s_1, \bm p's_1')$, $M(\bm ks, \bm k's' \to \bm ps_1, \bm p's_1')$, are calculated following the procedure outlined in Ref.~\cite{glazov2009}. They can be efficiently expressed in the terms of a function $\hat \mu(z, \bm k \to \bm p, \hat {\bm \sigma})$ which reads
\begin{widetext}
\begin{multline}
 \label{muf}
\hat \mu(z, \bm k \to \bm p, \hat {\bm \sigma}) = \varphi^2(z) + \mathrm i \xi \left\{ \hat \sigma_z [\bm p \times \bm k] \varphi^2(z) - \mathrm i [\hat {\bm \sigma} \times (\bm p + \bm k)]_z \varphi'(z) \varphi(z) + \right. \\
\left. \mathrm i \beta[\hat \sigma_x(p_x+k_x)-\hat \sigma_y(p_y+k_y)][\varphi'(z)]^2-\right. 
\left. \mathrm i \beta[\hat \sigma_x p_yk_y(p_x+k_x)-\hat \sigma_yp_xk_x(p_y+k_y)]\varphi^2(z) \right\},
\end{multline}
\end{widetext}
where the parameter $\xi$:
\begin{equation}
\label{xi}
\xi = 2AB+B^2 = -\frac{P^2}{3}\frac{\Delta(2E_g+\Delta)}{E_g^2(E_g+\Delta)^2},
\end{equation}
characterizes the strength of the spin-orbit interaction. Its values for some semiconductors are given in Table~\ref{table:xi}.
Here we neglected the corrections to the first three terms in Eq.~\eqref{muf} caused by the  electron energy spectrum non-parabolicity. The terms proportional to $\beta$ arise due to the bulk inversion asymmetry. 

\begin{table}[htbp]
\caption{Spin orbit interaction strength $\xi$~\cite{lawaetz71}.}
\begin{center} 
\begin{tabular}{cc}
\hline 
Material &  $\xi$ (\AA$^2$) \\
\hline 
GaAs & 5 \\
InAs & 100 \\
InSb & 500 \\
CdSe & 3 \\
CdTe & 5 \\
ZnTe & 2 \\
\hline
\end{tabular}
\label{table:xi}
\end{center}
\end{table}

Finally
\begin{multline}
 \label{Coulomb0}
M(\bm ks, \bm k's' \to \bm ps_1, \bm p's_1') =  \\
\delta_{\bm k+\bm k',\bm p + \bm p'}   
\int \mathrm dz_1 \mathrm dz_2 V(\bm p - \bm k,z_1-z_2) \times \\  \langle \chi_{s_1}\chi_{s_1'} |\hat \mu(z_1, \bm k \to \bm p, \hat{\bm \sigma}^{(1)})  \hat \mu(z_2, \bm k' \to \bm p', \hat{\bm \sigma}^{(2)}) | \chi_{s}\chi_{s'}\rangle,
\end{multline}
where $V(\bm q, z) = \Xi^{-1}\int V(\bm r) \mathrm e^{-\mathrm i \bm q \bm \rho} \mathrm d\bm \rho$ with $\Xi$ being the normalization area, $\bm q = \bm p - \bm k$ is the transferred wave vector, $\hat{\bm \sigma}^{(1)}$ and $\hat{\bm \sigma}^{(2)}$ are the spin operators of the first and second electrons (acting on spinors $|\chi_s\rangle$,$|\chi_{s_1}\rangle$ and on  $|\chi_{s'}\rangle$,$|\chi_{s_1'}\rangle$, respectively). Note, that the matrix element Eq.~\eqref{Coulomb0} is derived for not (anti)symmetrized wave functions, c.f.~\cite{ll4_eng}.

One can represent the matrix element Eq.~\eqref{Coulomb0} as a sum of spin-orbit independent contribution, $M^{(0)}$, the contributions linear in spin operators $\hat{\bm \sigma}^{(1)}$ and $\hat{\bm \sigma}^{(2)}$, $M^{(1)}$, and the quadratic in spin operator terms, $M^{(2)}$.

In what follows we consider a simple model of the quantum well of width $d$ with infinite barriers. The electron envelope function for the lowest size-quantization subband is taken in form
\begin{equation}
 \label{wavefunction}
\varphi(z) = \sqrt{\frac{2}{d}} \left[\cos{\left(\frac{\pi z}{d}\right)} + \alpha \sin{\left(\frac{2\pi z}{d}\right)}\right], \quad -d/2\leqslant z\leqslant d/2
\end{equation}
The possible quantum well heteropotential asymmetry (i.e. structural inversion asymmetry)  is taken into account by the second term in the square brackets, it as assumed that $\alpha \ll 1$. The effective matrix elements can be expressed in the terms of the form-factors~\cite{glazov2009}
\[
F_{ij}^{kl}(q) = 
\int \mathrm dz_1\mathrm dz_2 e^{-q|z_1 - z_2|} [\varphi(z_{1})]^i [\varphi(z_2)]^j [\varphi'(z_1)]^k[\varphi'(z_2)]^l,
\]
where $i, \ldots, l$ denote the powers. Hereafter it is assumed that $qd \ll 1$ which allows to consider only lowest size quantization subband of the quantum well and disregard the spread of the wave function along the growth direction.

The spin-orbit independent contribution is given by the Fourier transform of quasi two-dimensional Coulomb potential~\cite{glazov04a}
\begin{multline}
\label{direct:spinless}
M^{(0)}(\bm ks, \bm k's' \to \bm ps_1, \bm p's_1') = \\ \frac{2\pi e^2}{\Xi \varkappa q}  \delta_{\bm k+\bm k',\bm p + \bm p'} \delta_{s,s_1}\delta_{s',s_1'} F_{22}^{00}(q),
\end{multline}
with $F_{22}^{00}(q) =1$ at $qd \ll 1$.

The linear in spin operator contributions are responsible for the asymmetric scattering. They are proportional to $\xi$ and read
\begin{widetext}
\begin{multline}
 \label{Coulomb1}
M^{(1)}(\bm ks, \bm k's' \to \bm ps_1, \bm p's_1') = \xi \frac{2\pi e^2}{\Xi \varkappa q} \delta_{\bm k+\bm k',\bm p + \bm p'}
\langle  \chi_{s_1}\chi_{s_1'} |  
 [\hat{\bm \sigma}^{(1)} \times (\bm p+\bm k)]_z F_{12}^{10}(q)  +  [\hat{\bm \sigma}^{(2)} \times (\bm p'+\bm k')]_z F_{21}^{01}(q) -
\\ 
\beta[\hat{\sigma}^{(1)}_x(p_x+k_x) - \hat{\sigma}^{(1)}_y(p_y+k_y)] F_{02}^{20}(q)-\beta[\hat{\sigma}^{(2)}_x(p_x'+k_x') - \hat{\sigma}^{(2)}_y(p_y'+k_y')] F_{20}^{02}(q) +
\\
\mathrm i \hat{\sigma}^{(1)}_z[\bm p \times \bm k]_zF_{22}^{00}(q)  + \mathrm i \hat{\sigma}^{(2)}_z[\bm p' \times \bm k']_z F_{22}^{00}(q)\left.\left. \right]\right\}| \chi_{s}\chi_{s'}\rangle.
\end{multline}
\end{widetext}
Here we retained only linear and quadratic in wave vectors terms. The first line in Eq.~\eqref{Coulomb1} describe Rashba-like contributions to electron-electron scattering similar to the structural inversion asymmetry terms in electron impurity or electron phonon scattering~\cite{averkiev02,tarasenko05eng}. However, the form-factors $F_{12}^{10}=F_{21}^{01}\equiv 0$ in our model and structural inversion asymmetry terms vanish.\footnote{These terms can be interpreted as Rashba effect for a given electron caused by the other one. In our model they vanish because that the total $z$-component of electric field induced by some charge distribution acting on the same charge distribution is zero. [E.Ya. Sherman, private communication]} The second line of Eq.~\eqref{Coulomb1} describes bulk inversion asymmetry terms ($\propto \beta$). The corresponding form-factors are $F_{02}^{20}(q) = F_{20}^{02}(q) = {\pi^2}/{d^2}$. Finally, the line line describes Mott (skew) scattering. These linear in spin terms can cause the spin currents generation due to electron-electron interaction~\cite{Badalyan-2009}.

The quadratic in spin operator terms describe spin-spin interaction and can be written as:
\begin{widetext}
\begin{multline}
 \label{Coulomb2}
M^{(2)}(\bm ks, \bm k's' \to \bm ps_1, \bm p's_1') = \xi^2 \frac{2\pi e^2}{\Xi \varkappa q} \delta_{\bm k+\bm k',\bm p + \bm p'}\langle \chi_{s_1}\chi_{s_1'} |  \left\{  [\hat{\bm \sigma}^{(1)} \times (\bm p+\bm k)]_z[\hat{\bm \sigma}^{(2)} \times (\bm p'+\bm k')]_z F_{11}^{11}(q) + \right.\\
 \beta^2[\hat{\sigma}^{(1)}_x(p_x+k_x) - \hat{\sigma}^{(1)}_y(p_y+k_y)] [\hat{\sigma}^{(2)}_x(p_x'+k_x') - \hat{\sigma}^{(2)}_y(p_y'+k_y')] F_{00}^{22}(q)-
\\ \beta[\hat{\sigma}^{(1)}_x(p_x+k_x) - \hat{\sigma}^{(1)}_y(p_y+k_y)] [\hat{\bm \sigma}^{(2)} \times (\bm p'+\bm k')]_z
F_{01}^{21}(q) - \\
\beta [\hat{\bm \sigma}^{(1)} \times (\bm p+\bm k)]_z[\hat{\sigma}^{(2)}_x(p_x'+k_x') - \hat{\sigma}^{(2)}_y(p_y'+k_y')] F_{10}^{12}(q) -
  \\
\left.\left.([\bm p \times \bm k]\hat{\bm \sigma}^{(1)} )([\bm p' \times \bm k']\hat{\bm \sigma}^{(2)}) F_{22}^{00}(q) \right]\right\}| \chi_{s}\chi_{s'}\rangle.
\end{multline}
\end{widetext}
Here we disregarded the terms with odd powers of the wave vectors and the terms with the powers of the wave vectors higher than $4$: odd terms do not result in a two-electron states fine structure, and higher order terms do not change the results qualitatively. The terms in the first and last lines of Eq.~\eqref{Coulomb2} were derived in Ref.~\cite{glazov2009}, the form-factor $F_{11}^{11}(q)=3q/(4d)$. The allowance for the bulk inversion asymmetry results in the terms proportional to $\beta$ and $\beta^2$. The form-factor at the $\beta^2$ terms is $F_{00}^{22}(q) = \pi^4/d^4$. Note, that the terms linear in $\beta$ arise due to an interplay of bulk and structural inversion asymmetry, corresponding form-factors $F_{10}^{12}(q) = F_{01}^{21}(q) = -128\alpha q/(15d^2)$. These form-factors are non-zero in symmetric quantum wells.

\subsection{Allowance for the heavy-light hole splitting}

So far, we assumed that the valence band $\Gamma_8^{\rm v}$ states are degenerate at $\bm k =0$. The deformation of a cubic semiconductor along the $z$ axis lifts this degeneracy, see Fig.~\ref{fig:kane14}(b). In quantum wells the size quantization removes the degeneracy of the states with the total momentum projection $\pm 3/2$ and $\pm 1/2$ on the growth axis. Similar model can be applied to a certain extent to wurzite semiconductors with $z$ being the wurzite axis~\cite{birivch75}.

In order to analyze the effect of the valence band splitting on the spin-orbit contributions to electron-electron interaction Hamiltonian we disregard the bulk inversion asymmetry (i.e. put $\beta=0$). We denote as $E_g$ the distance between $\Gamma_6^{\rm c}$ band and a light-hole band ($z$-component of the hole angular momentum being $\pm 1/2$) and as $\tilde E_g$ the distance between $\Gamma_6^{\rm c}$ band and a heavy-hole band ($z$-component of the angular momentum being $\pm 3/2$). Let $\Delta$ be the splitting between the light-hole band and the spin-orbit split-off band $\Gamma_7^{\rm v}$, see Fig.~\ref{fig:kane14}(b). We assume that the heavy-light hole splitting $\tilde E_g - E_g$ is small as compared with $\Delta$. Under this assumption function $\hat\mu$ in Eq.~\eqref{Coulomb0} takes the form
\begin{multline}
 \label{muf1}
\hat \mu(z, \bm k \to \bm p, \hat {\bm \sigma}) = \varphi^2(z) + \\ \mathrm i (\xi+ \tilde\xi)  \hat \sigma_z [\bm p \times \bm k] \varphi^2(z) + \xi [\hat {\bm \sigma} \times (\bm p + \bm k)]_z \varphi'(z) \varphi(z) ,
\end{multline}
where $\xi$ is given by Eq.~\eqref{xi} (with the notations above), and 
$\tilde \xi = -P^2(E_g^2-\tilde E_g^2)/({2E_g^2\tilde E_g^2})$. The splitting of the hole states results in the change of the coefficient at the term, proportional to $\hat\sigma_z$.

\subsection{Short-range electron-electron interaction}

The method developed in Ref.~\cite{glazov2009} and extended here allows to determine the long-range contributions to the electron-electron interaction which are caused by the Fourier components of the Coulomb potential with the transfered wave vector much smaller as compared with the inverse lattice constant. Within the framework of the effective mass method other Fourier components of the Coulomb potential form a short-range electron-electron interaction:
\begin{equation}
\label{short:range}
\mathcal H^{\rm short}(\bm r, \bm r') =  \gamma_{ij} \ \hat{ \sigma}^{(1)}_i \hat{\sigma}^{(2)}_j  \ \delta(\bm r - \bm r'),
\end{equation}
where the summation over the repeated Cartesian subscripts $i,j=x,y,z$ is assumed. Here the spin-independent contributions are ignored and the non-zero components of the tensor $\gamma_{ij}$ are determined by the point symmetry of the considered system. Evaluation of the tensor $\gamma_{ij}$ requires a fully microscopic calculation.

In bulk cubic semiconductors the tensor $\gamma$ reduces to the scalar and the short-range interaction Eq.~\eqref{short:range} simply adds to the Coulomb exchange interaction caused by the (anti)symmetry of the wave functions and, hence, the short-range contribution can be disregarded. In bulk deformed semiconductors and in wurzite systems there are two independent components of tensor $\gamma$: $\gamma_{zz}$ and $\gamma_{\perp}\equiv\gamma_{xx}=\gamma_{yy}$.

The effective short-range interaction Hamiltonian in low-dimensional systems: quantum wells, wires and QDs can be obtained by averaging of Eq.~\eqref{short:range} with the  size-quantization wave functions of interacting electrons. It is worth noting that the additional terms may arise due to the interfaces of the low-dimensional systems~\cite{aleiner92eng}.

\section{Two-electron triplet state fine structure}

Here we apply the developed formalism to calculate the fine structure of two electrons localized either in a QD or in a lateral double QD. In what follows we consider triplet states which are described by antisymmetric combination of single electron in-plane envelopes $\psi_1(\bm \rho_1)$, $\psi_2(\bm \rho_2)$:
\begin{equation}
\label{triplet}
\Psi(\bm \rho_1,\bm \rho_2) = \mathcal N \left[\psi_1(\bm \rho_1)\psi_2(\bm \rho_2) - \psi_1(\bm \rho_2)\psi_2(\bm \rho_1) \right],
\end{equation}
where $\mathcal N$ is the normalization constant.
The singlet-triplet splitting caused by the Coulomb exchange interaction $2U_e$, where
\begin{equation}
\label{exchange}
U_e = \frac{2\mathcal Ne^2}{\varkappa} \int \frac{\mathrm d \bm \rho_1 \mathrm d\bm \rho_2}{|\bm \rho_1 - \bm \rho_2|} \psi_1(\bm \rho_1) \psi_2(\bm \rho_1)\psi_2(\bm \rho_2)\psi_1(\bm \rho_2),
\end{equation}
is assumed to exceed by far the fine structure splittings of the triplet states.

In the absence of the spin-orbit interaction the triplet state Eq.~\eqref{triplet} is three-fold spin degenerate with respect to the $z$-component of the total electron spin: $m_z=0$, $1$ and $-1$. The spin-orbit interaction combined with the Coulomb interaction between the charge carriers may lift this degeneracy. In the QD systems described by $C_{2\rm v}$ point symmetry group with the main in-plane axes $x'\parallel [1\bar 10]$, $y'\parallel [110]$ and axis $z'\parallel [001]$, the effective Hamiltonian acting in the basis of the three states $|m_z\rangle$ which describes their fine structure can be represented via the operators of the total momentum $1$, $\hat S_i$ ($i=x',y',z'$) as
\begin{equation}
 \label{H:sym:ii}
\hat{\mathcal H} =  \mathcal A \hat S_{x'}^2 + \mathcal B \hat S_{y'}^2 - (\mathcal A + \mathcal B) \hat S_{z'}^2,
\end{equation}
by means of two independent constants $\mathcal A$ and $\mathcal B$. The term proportional to $\hat S_z^2$ is added in order to eliminate the irrelevant total energy shift of the triplet. 

There are three non-degenerate eigenstates of the Hamiltonian~\eqref{H:sym:ii}~\cite{glazov2009}: one state with energy $E_0=\mathcal A+\mathcal B$ characterized by the total spin projection on the growth axis being $0$ and ``linearly-polarized'' combinations:
\begin{equation}
\label{x1y1}
|x'\rangle = \frac{1}{\sqrt{2}}(|+1\rangle+|-1\rangle), \quad |y'\rangle = -\frac{\mathrm i}{\sqrt{2}}(|+1\rangle-|-1\rangle),
\end{equation}
with energies $\mathcal E_{x'} = - \mathcal B$ and $\mathcal E_{y'} = -\mathcal A$, respectively, similar to those of heavy-hole exciton in an anisotropic quantum disk~\cite{ivchenko05a,goupalov98,glazov2007a}. The values of the constants $\mathcal A$ and $\mathcal B$ depend on the wave function shapes and on the spin splitting parameters $\xi$, $\tilde \xi$ and $\beta$.

\subsection{Single quantum dot}

First we consider a single parabolic QD. As an example we take the lowest exited $SP$ triplet state where one electron occupies ground state $S$-shell orbital and the other one occupies the lowest excited $P$-shell state:
\begin{equation}
\label{qd:states}
\psi_1(\bm \rho) = \frac{1}{\sqrt{2\pi a^2}} e^{-\rho^2/4a^2} ,\quad
\psi_2(\bm \rho) = \frac{x'}{a} \psi_1(\bm \rho), 
\end{equation}
where $a$ is an effective disk radius. It is assumed that the QD is slightly elongated along $x'$ axis so that $P_{x'}$ orbital [given by Eq.~\eqref{qd:states}] is lower in energy as compared with the $P_{y'}$ orbital with the wave function $(y'/a)\psi_1(\bm \rho)$. The deformation of the wave functions due to the ellipticity of the QD is neglected~\cite{glazov2009}. The singlet-triplet splitting in such a QD equals to $2U_e = \sqrt{\pi} e^2/(4\varkappa a)$.

\begin{figure}
\includegraphics[width=\linewidth]{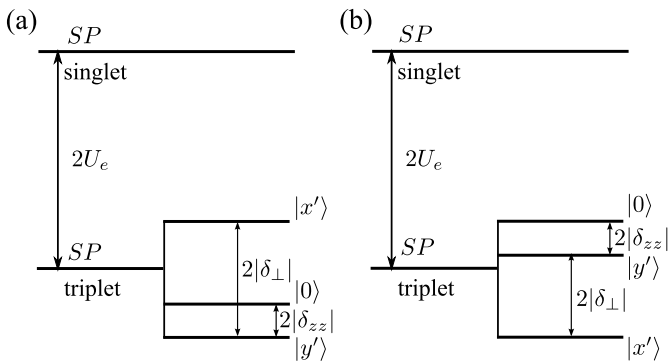}
\caption{Schematic illustration of the two-electron energy levels fine structure in an anisotropic quantum disk. Here $|0\rangle$, $|x'\rangle$ and $|y'\rangle$ denote spin states (i.e. those with the total spin $z$ projection $m_z$ being $0$, and two linear combinations of $m_z=\pm 1$ states). The singlet state lies above the triplet states and is not affected by the spin-orbit interaction. Panel (a): negligible structural inversion asymmetry ($\delta_\perp<0$), panel (b) comparable bulk and structural inversion asymmetry ($\delta_\perp>0$). Splittings are shown not to scale.}
\label{fig:aniso} 
\end{figure}

The constants $\mathcal A$ and $\mathcal B$ describing the fine structure of two-electron triplet state can be conveniently expressed in the terms of new parameters $\delta_{\perp}$ and $\delta_{zz}$ describing the splittings between the linearly polarized states and between the state with $m_z=0$ and one of the linearly-polarized states, namely, $|y'\rangle$, respectively, as it is shown in Fig.~\ref{fig:aniso}:
\begin{equation}
 \label{Delta:qd:1}
\mathcal A = - \frac{2}{3}(\delta_\perp + \delta_{zz}), \quad \mathcal B = \frac{2}{3}(2\delta_\perp - \delta_{zz}).
\end{equation}
The calculation of matrix elements Eq.~\eqref{Coulomb2} with the antisymmetrized wave function \eqref{triplet} neglecting heavy-light hole splitting shows that the direct contribution vanishes (because $M^{(2)}$ is represented as a product of operators acting on each electron states) and the exchange contribution yields:
\begin{equation}
\label{deltas}
\delta_{zz} = - \xi^2 \frac{\sqrt{\pi}e^2}{32 \varkappa a^5}, \quad \delta_{\perp} =  -\xi^2\frac{e^2}{\varkappa d}\left[\frac{3}{8 a^4}+ \frac{\pi^{9/2} \beta^2 }{2d^3a^3}-\frac{128 \alpha\beta}{15da^4}\right].
\end{equation}
There are three contributions to $\delta_{\perp}$: the first one calculated in Ref.~\cite{glazov2009} is not related with the lack of an inversion center, the second one results from the bulk inversion asymmetry and the third one is a result of an interplay between bulk and structural inversion asymmetry. Two last terms in Eq.~\eqref{deltas} are zero if bulk inversion asymmetry is disregarded ($\beta=0$), moreover, the third contribution to $\delta_{\perp}$ vanishes at $\alpha=0$, i.e. if the heteropotential has an inversion center.

It follows from Eqs.~\eqref{Delta:qd:1}, \eqref{deltas} that the spin degeneracy of the triplet two-electron states is fully lifted in anisotropic QDs by the combination of the exchange and spin-orbit interactions (spin-orbit exchange). In perfectly isotropic QDs the situation may be different~\cite{glazov2009}.

It is worth noting, that depending on the values of $\alpha$ and $\beta$ the sign of $\delta_{\perp}$ can be arbitrary. The negative sign of $\delta_\perp$ corresponds to the small structural inversion asymmetry, while positive sign of $\delta_\perp$ corresponds to the comparable bulk and structural inversion asymmetry, $\alpha\beta>0$, with the last term in Eq.~\eqref{deltas} being dominant. It results in a different order of levels: if $\delta_\perp<0$ when the state with $m_z=0$ lies between the states $|x'\rangle$ and $|y'\rangle$~Fig.~\ref{fig:aniso}(a), otherwise it lies above the doublet $|x'\rangle$ and $|y'\rangle$~Fig.~\ref{fig:aniso}(b). Since $\delta_{zz}<0$ the state $|0\rangle$ always lies above the state $|y'\rangle$.

The allowance for the heavy-light hole splittings results in the replacement of $\xi$ by $\xi+\tilde{\xi}$ in the expression for $\delta_{zz}$, Eq.~\eqref{deltas}. If $\tilde E_g< E_g$, i.e. the valence band top is determined by the heavy holes, see Fig.~\ref{fig:kane14}(b), then $|\delta_{zz}|$ increases.

\subsection{Lateral double quantum dot}

The fine structure of triplet two-electron states in lateral double QDs is qualitatively similar to the case of $SP$ electron states in a single QD considered above.

We assume that the electron states in each dot can be described by Gaussian wave functions~\cite{badescu:161304}
\begin{equation}
\label{dd:states}
 \psi_1(\bm \rho) = \dfrac{1}{\sqrt{2\pi} a} \mathrm e^{-\rho^2/4a^2}, \quad
 \psi_2(\bm \rho) = \dfrac{1}{\sqrt{2\pi} a} \mathrm e^{-(\bm \rho - \bm L)^2/4a^2},
\end{equation}
where $\bm L$ is a vector connecting QD centers. The overlap between these states $\sim \exp{(-L^2/4a^2)}$ is supposed to be small so that the triplet state can be written in the form of Eq.~\eqref{triplet}. The normalization constant $\mathcal N =[2- 2\exp{(-L^2/4a^2)}]^{-2}\approx 1/\sqrt{2}$. The Coulomb exchange interaction constant $U_e$ given by Eq.~\eqref{exchange} equals approximately to\footnote{It was obtained by fitting the results of numerical integration.} $0.89\exp{(-L^2/4a^2)} e^2/(\varkappa a)$ and is assumed to exceed the fine structure splittings of the triplet state.

In what follows we disregard the bulk inversion asymmetry $\beta=0$ and splitting of the heavy and light hole bands. It is convenient to express the constants $\mathcal A$ and $\mathcal B$ in effective Hamiltonian~\eqref{H:sym:ii} via the parameters $\delta_{\perp}$ and $\delta_{zz}$ by means of Eq.~\eqref{Delta:qd:1}. The calculation shows that\footnote{Value of $\delta_{zz}$ was obtained by fitting the results of numerical integration.}
\begin{equation}
\label{deltas:double}
\delta_{zz} = -0.014\frac{e^2\xi^2}{a^5\varkappa} \frac{L^2}{a^2} \mathrm e^{-\frac{L^2}{4a^2}},
\quad 
\delta_\perp = -\frac{3e^2\xi^2}{32d a^4\varkappa} \frac{L^2}{a^2} \mathrm e^{-\frac{L^2}{4a^2}}.
\end{equation}
The qualitative level arrangement is similar to the one shown for a single QD in Fig.~\ref{fig:aniso}(a). The allowance for the bulk and structural inversion asymmetries may transform the level arrangement to the one shown in Fig.~\ref{fig:aniso}(b). In the latter case the splittings and eigenspin states are strongly sensitive to the orientation of the vector $\bm L$ relative to the in-plane axes. 

It is instructive to analyze an interplay of the electron-electron spin-orbit exchange interaction and the spin splitting of the conduction band caused by bulk and structural inversion asymmetry. In the relevant case where the interdot distance $L$ exceeds by far the QD size $a$ the spin-orbit splitting of the conduction band can be eliminated in the lowest order by a unitary transformation which takes into account the rotation of electron spin during the electron tunneling~\cite{kkavokin01,gangadharaiah:156402}. Therefore, the exchange interaction between the electrons expressed in terms of the transformed spins has the same form as without the spin-orbit splitting, Eq.~\eqref{H:sym:ii}. Therefore, the level splittings are the same as without the conduction band spin splitting, but the eigenstates are different because they correspond to the rotated spins. The higher order effects of the conduction band spin-orbit splitting may give rise to an additional contribution to $\delta_{zz}$~\cite{gangadharaiah:156402}.

Finally, we compare the ratio of the spin-splitting $|\delta_{zz}|$ to the exchange Couloumb interaction energy for the $SP$ state in a single QD and for the double QD:
\[
 \left| \frac{\delta_{zz}}{U_e}\right| \approx \dfrac{\xi^2}{a^4}\times \left\{
 \begin{array}{cc}
 0.25, & \mbox{single dot},\\
 0.016 \dfrac{L^2}{a^2}, & \mbox{double dot}.
 \end{array}
 \right.
\]
Therefore the relative role of the spin-orbit effects in double dots is higher: although the absolute values of the fine structure constants decrease with an increase of the dot separation $L$, the ratio of the spin splittings of the triplet and the exchange interaction constant increases.

\section{Photoluminescence spectra}

In this section we calculate the photoluminescence spectra of a single QD in two particular situations where the fine structure of triplet two-electron states is manifested: first, we consider the emission of a ``hot'' (excited) trion and, second, we consider the emission of a doubly charged exciton.

\subsection{Hot trion}

The hot trion is formed from a hole and two electrons: one in the ground and the second one in the excited state in the QD. In what follows we assume that two electrons occupy $S$ and $P$ orbitals and constitute the orbital state described by the antisymmetric wave function, Eq.~\eqref{triplet}. Although this state is excited, its relaxation to the ground, singlet trion state where two electrons occupy the same orbital can be strongly suppressed~\cite{cortez02}. A fine structure of the excited trion is determined by an interplay of the spin-orbit interaction and the electron-hole exchange interaction.

\begin{figure}[hptb]
 \includegraphics[width=\linewidth]{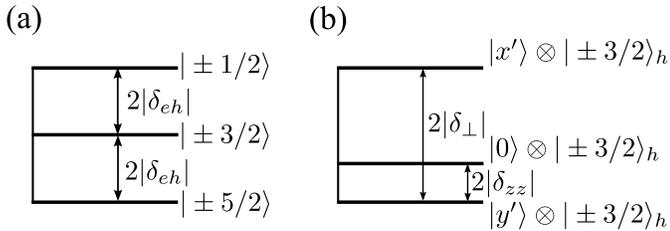}
\caption{Hot trion fine structure. (a) Dominant electron-hole exchange interaction: $|\pm 1/2\rangle$, $|\pm 3/2\rangle$, and $|\pm 5/2\rangle$ denote the total spin (electron and hole) projection on the growth axis. (b) Dominant electron-electron spin-orbit exchange interaction: $|x'\rangle$, $|y'\rangle$ and $|0\rangle$ denote the two-electron spin states, Eq.~\eqref{x1y1}, $|\pm 3/2\rangle_h$ is the hole spin state.}
\label{fig:hot}
\end{figure}

\begin{figure}[hptb]
 \includegraphics[width=0.95\linewidth]{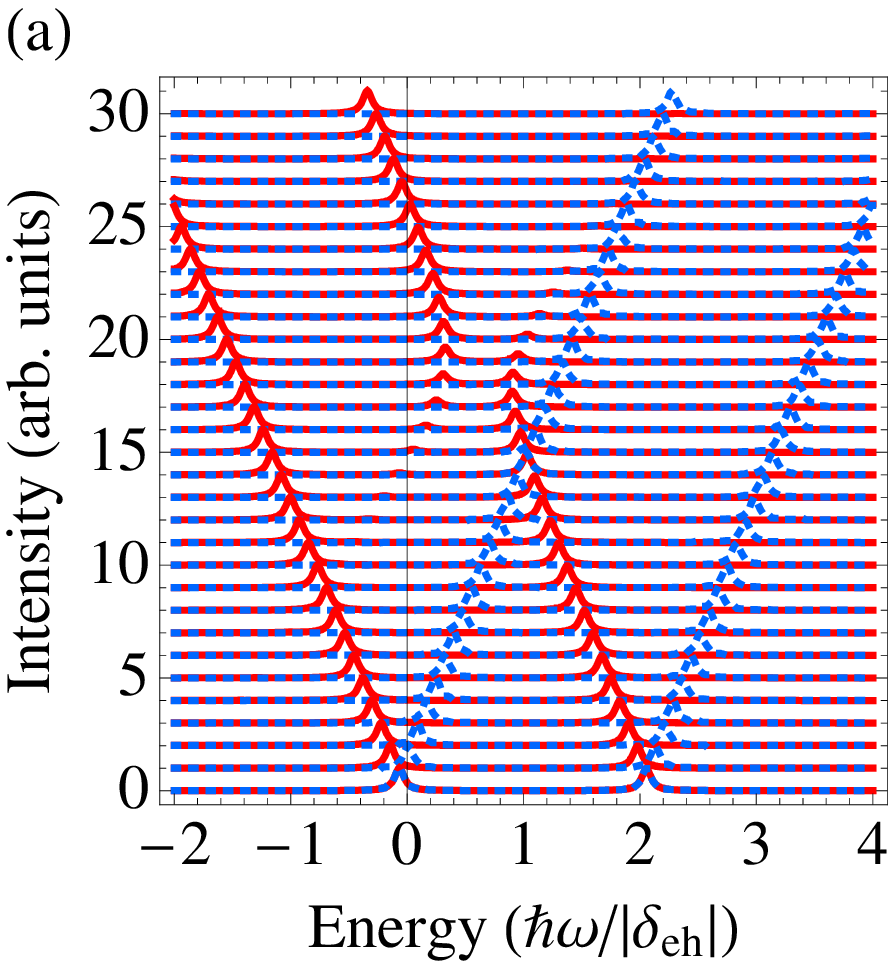}
 \includegraphics[width=0.95\linewidth]{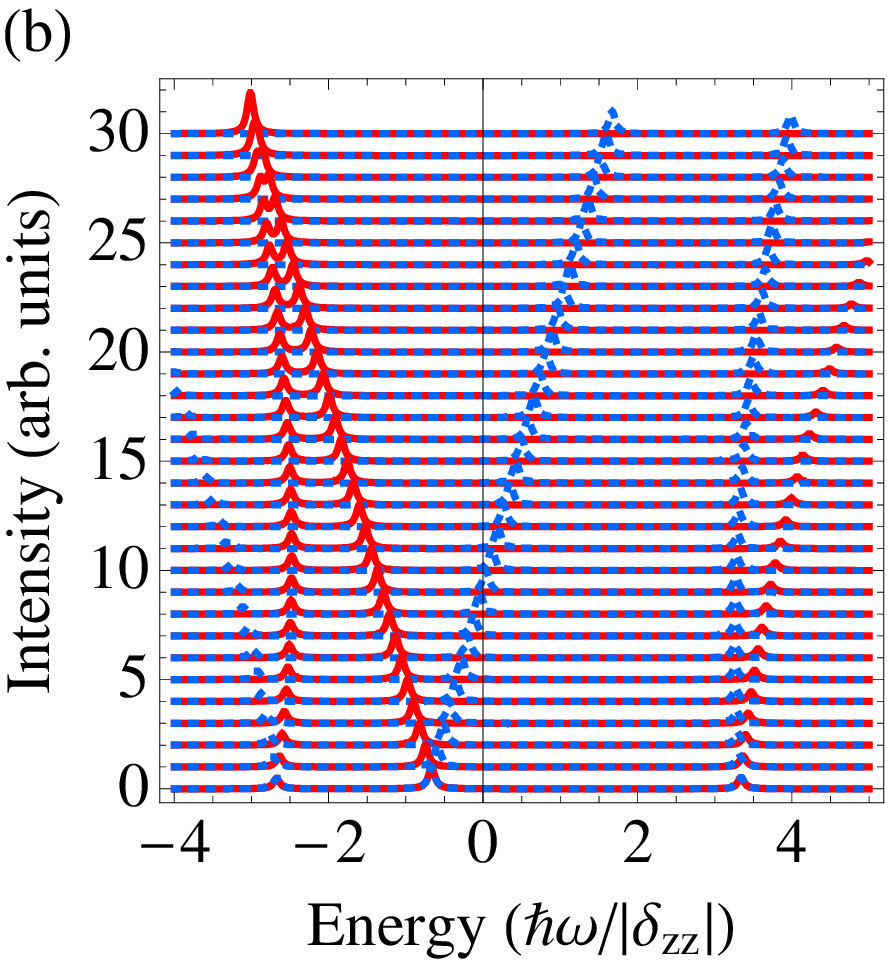}
\caption{Photoluminescence spectra of hot trion. Different lines (offset along the vertical axis for clarity) correspond to different values of an external magnetic field applied along $z$ axis. Red/solid curve corresponds to the emission in $\sigma^+$ polarization, blue/dotted curve corresponds to the emission in $\sigma^-$ polarization. Electron and hole effective $g$-factors are $g_e>0$, $3g_h=-0.42g_e$. Panel (a): relatively small electron-electron spin orbit exchange interaction, $\delta_{zz}=0.1\delta_{eh}$, $\delta_{\perp}=3\delta_{zz}$, $\delta_{eh}<0$ and $|\delta_{eh}|$ is used as a unit of energy. The energy is referred from the position of the state with the total spin $z$-component being $\pm 3/2$ calculated neglecting electron-electron spin orbit coupling. Panel (b): relatively large electron-electron spin orbit exchange interaction,  $\delta_{\perp}=3\delta_{zz}$, $\delta_{zz}<0$, $\delta_{eh}=0.1\delta_{zz}$ and $|\delta_{zz}|$ is used as a unit of energy. The energy is referred from the center of a two-electron triplet states calculated neglecting electron-hole exchange interaction. The lines are Lorentzian broadened with the width $\Gamma=0.05|\delta_{zz}|$.}
\label{fig:hotpl}
\end{figure}

The short-range part of the exchange interaction Hamiltonian between the hole and two electrons in the $SP$ excited state is given by
\begin{equation}
\hat{\mathcal H}_{eh} = \frac{2}{3}\delta_{eh}[\hat{\sigma}_z^{(1)} + \hat{\sigma}_z^{(2)}] \hat{J}_z \ ,
\end{equation}
where $\hat J_z$ is the heavy-hole angular momentum projection operator and $\delta_{eh}$ ($\delta_{eh}<0$, as a rule) is a constant determining the splittings between the states with different $z$-components of the total spin (electron and hole), see Fig.~\ref{fig:hot}(a). Hence, if the electron-electron spin-orbit exchange interaction is negligible, the fine structure of a triplet hot trion consists of three two-fold degenerate sublevels split by $|2\delta_{eh}|$. Two of these sublevels $|\pm 1/2\rangle$ and $|\pm 3/2\rangle$ are optically active. This fine structure for a hot trion is widely accepted in literature~\cite{bracker08,kusraev08,cortez02} and indeed holds if $|\delta_{eh}|\gg |\delta_{\perp}|$ in Eq.~\eqref{Delta:qd:1}.

In the opposite limiting case, $|\delta_{eh}|\ll |\delta_{\perp}|$, the fine structure of an excited trion state is different, see Fig.~\ref{fig:hot}(b). There are three sublevels, each of those is two-fold degenerate with respect to the hole spin projection. The two-electron spin states are: one with $z$-projection being $0$ and two linear combinations, Eq.~\eqref{x1y1}. All these three states are optically active. At arbitrary relation between $\delta_{eh}$ and $\delta_{\perp}$ there are also three doubly degenerate levels which are, in general, active.

Hot trion photoluminescence spectra calculated in the case of relatively weak and relatively strong electron-electron spin-orbit exchange interaction are presented in Fig.~\ref{fig:hotpl}, panels (a) and (b), respectively. Different curves which are vertically shifted for the clarity correspond to different values of the magnetic field applied along the growth axis. The electron and hole effective $g$-factors which determine the energies of the spin-split states as $\pm g_e\mu_B B/2$ and $\pm 3g_h\mu_B B/2$ (where $\mu_B$ is Bohr magnetron and $B$ is the magnetic field $z$-component) are choosen to have different signs, $g_e>0$, $g_h<0$ which corresponds to CdSe/ZnSe/ZnMnSe QDs studied in Refs.~\cite{kul1:07}. Other parameters are presented in the caption to the Fig.~\ref{fig:hotpl}.

In the case of weak electron-electron spin-orbit exchange interaction, Fig.~\ref{fig:hotpl}(a), two doublets are clearly seen which correspond to the states $|\pm 1/2\rangle$ and $|\pm 3/2\rangle$. In $B=0$ the emission from the ``dark'' states $|\pm 5/2\rangle$ can not be seen from the figure because the bright states admixture due to electron-electron spin-orbit exchange is small. With an increase of the magnetic field an anticrossing is clearly seen, which is an evidence of the electron-electron spin-orbit interaction which intermixes dark $|5/2\rangle$ and bright $|-1/2\rangle$ states when they become resonant in a certain magnetic field. The observation of such an anticrossing may allow one to measure the electron-electron spin-orbit interaction constant $\delta_{\perp}$.

The emission spectra are completely rearranged in the case of dominant electron-electron spin-orbit exchange interaction, $|\delta_{\perp}|\gg |\delta_{eh}|$, Fig.~\ref{fig:hotpl}(b). In this case all the states are optically active: in zero magnetic field three lines are seen, corresponding to the eigenstates shown in Fig.~\ref{fig:hot}(b). In magnetic field applied along the growth direction six emission lines with different intensities are visible.

\subsection{Doubly charged exciton}

Here we consider the emission spectra of doubly charged exciton, $X^{2-}$, which is formed of three electrons and a hole. Two electrons occupy the ground $S$-shell state in a QD being in the spin singlet, and the third one occupies an excited $P$-orbital state. The hole occupies the ground state and radiatively recombines with one of $S$-shell electrons leaving remaining two electrons in the $SP$-orbital state.

\begin{figure}[hptb]
\includegraphics[width=\linewidth]{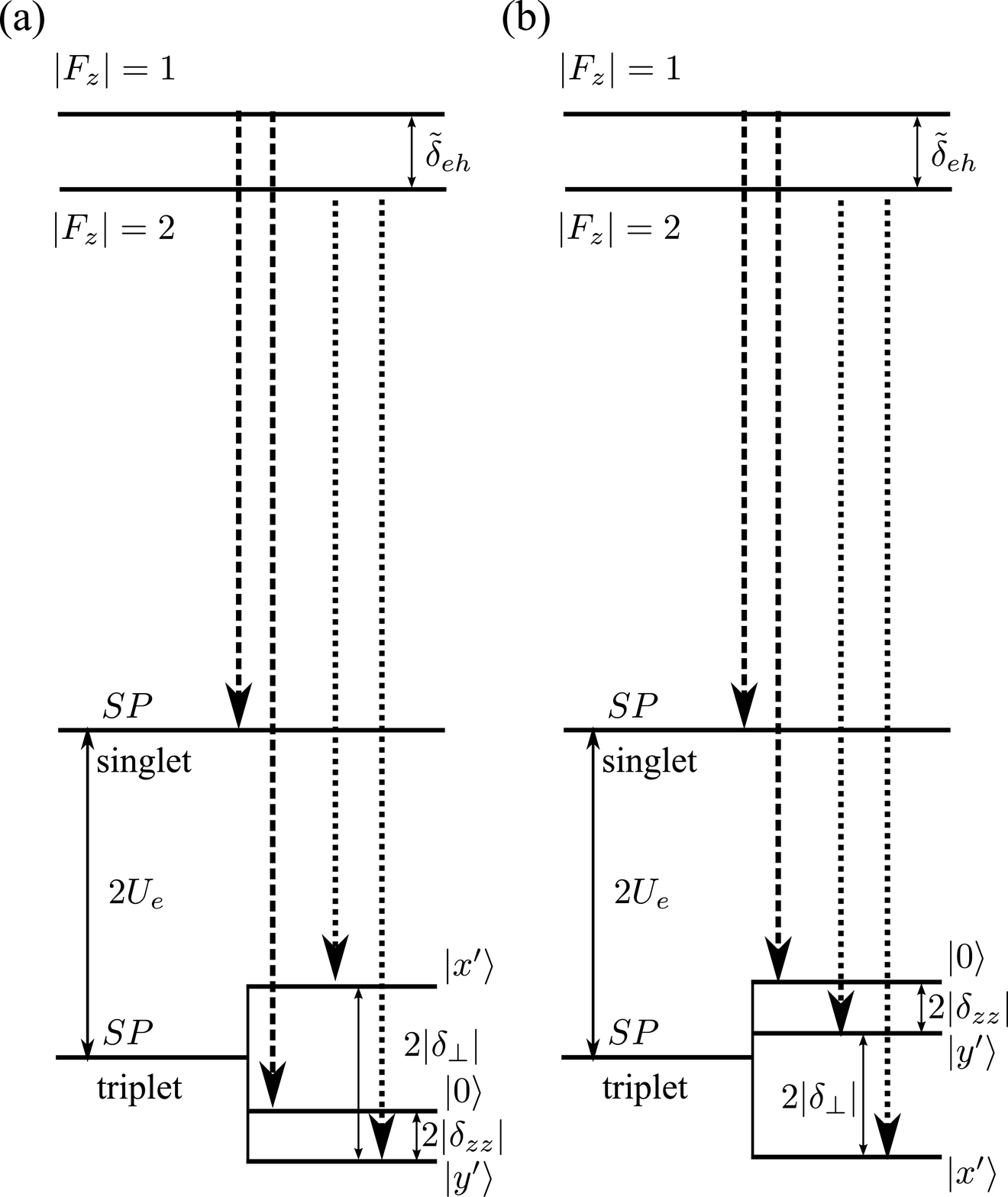}
\caption{Energy levels and optical transitions in $X^{2-}$ complex. States $|F_z|=1$, $|F_z|=2$ are the spin states of $X^{2-}$, states denoted as $SP$ singlet and triplet are the two-electron final states. Long dashed and short dashed vertical arrows denote the optical transitions from $|F_z|=1$ and $|F_z|=2$ states, respectively.  Panel (a): negligible structural inversion asymmetry ($\delta_\perp<0$), panel (b) comparable bulk and structural inversion asymmetry ($\delta_\perp>0$). Splittings are shown not to scale.}
\label{fig:x2-}
\end{figure}

The initial $X^{2-}$ state can be conveniently characterized by the $z$-component of the total spin of electrons and holes, $F_z$, which takes the following values $\pm 1$ and $\pm 2$, since two electrons form a spin singlet. The electron hole short-range exchange interaction splits the states with $|F_z|=1$ and $|F_z|=2$ by the value $\tilde{\delta}_{eh}$ similar to the splitting of the ground excitonic states into bright and dark doublets. In this case, however, both $|F_z|=1$ and $|F_z|=2$ states are optically active, the schemes of transitions for different level arrangements are depicted in Fig.~\ref{fig:x2-}.

\begin{figure*}[hptb]
 \includegraphics[width=0.45\textwidth]{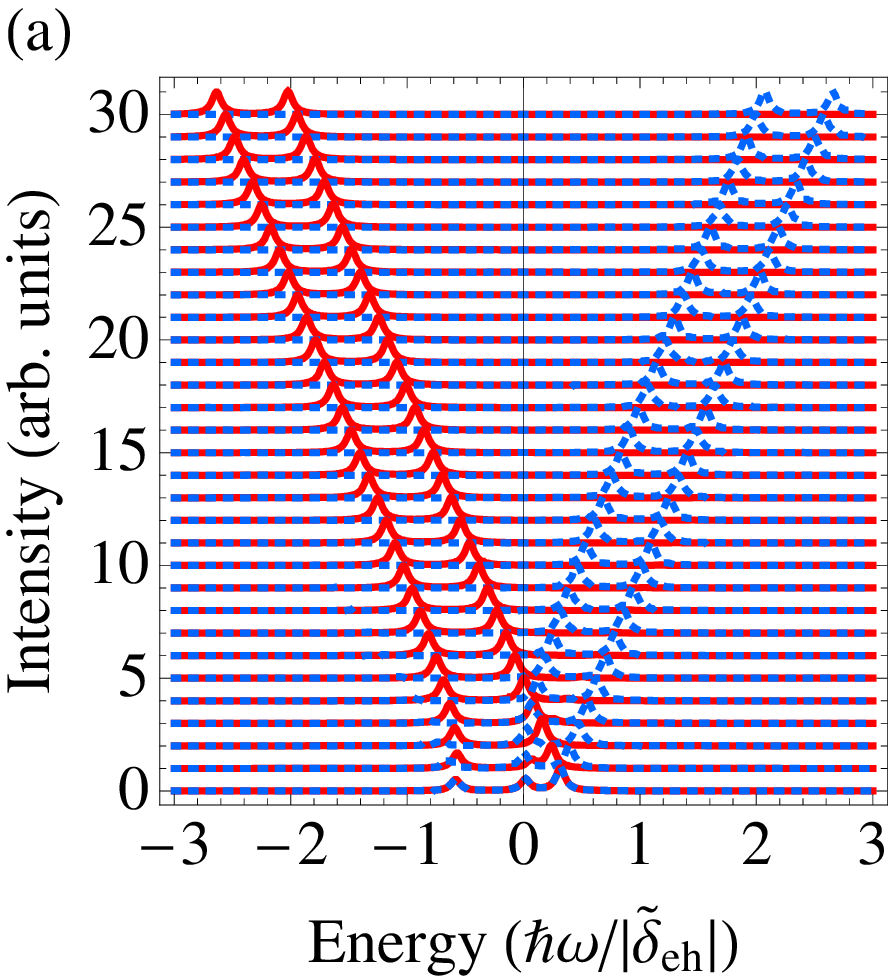}
 \includegraphics[width=0.45\textwidth]{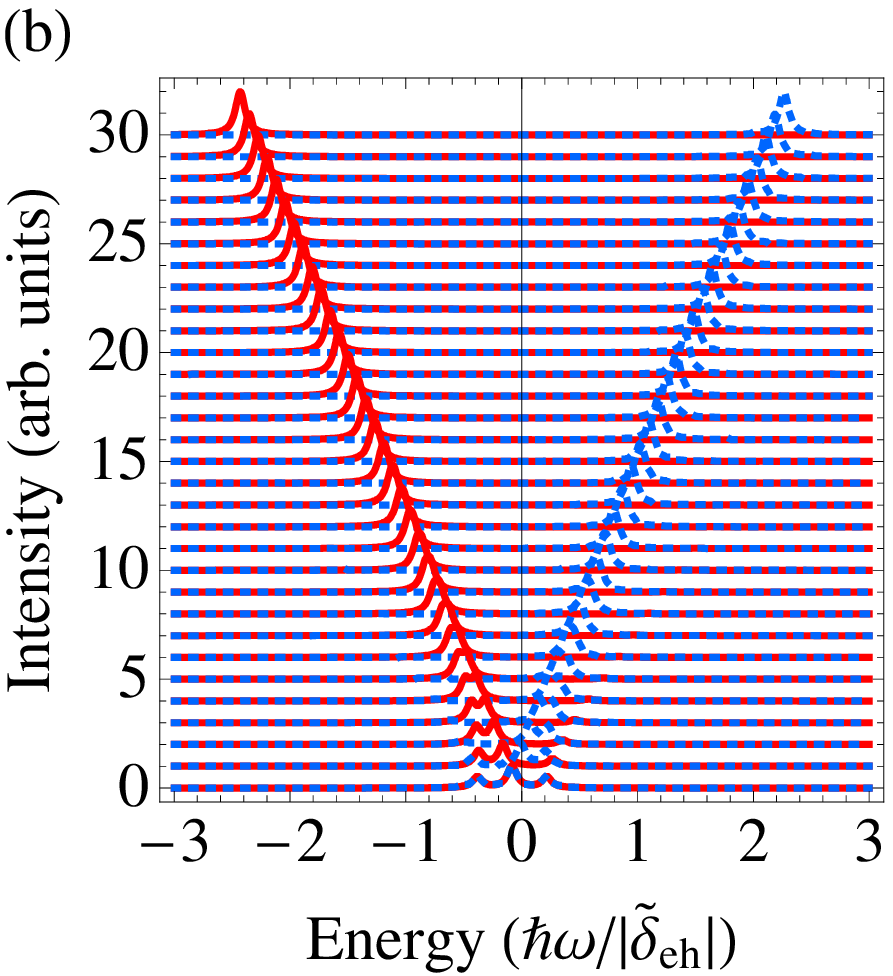}
\caption{Photoluminescence spectra of $X^{2-}$ complex. Different lines (offset along the vertical axis for clarity) correspond to different values of an external magnetic field applied along $z$ axis. Red/solid curve corresponds to the emission in $\sigma^+$ polarization, blue/dotted curve corresponds to the emission in $\sigma^-$ polarization. Electron and hole effective $g$ factors are $g_e>0$, $3g_h=-0.42g_e$. The energy unit is $|\tilde{\delta}_{eh}|$. Panel (a) corresponds to the two-electron level structure shown in Figs.~\ref{fig:aniso}(a), \ref{fig:x2-}(a) where structural inversion asymmetry is negligible: $\delta_{zz}=-0.1|\tilde{\delta}_{eh}|<0$, $\delta_{\perp}=3\delta_{zz}<0$.
Panel (b) corresponds to the two-electron level structure shown in Figs.~\ref{fig:aniso}(b), \ref{fig:x2-}(b) (comparable bulk and structural inversion asymmetry): $\delta_{zz}=-0.1|\tilde{\delta}_{eh}|<0$, $\delta_{\perp}=-3\delta_{zz}>0$.
 The lines are Lorentzian broadened with the width $\Gamma=0.05|\tilde{\delta}_{eh}|$. Transitions to singlet two-electron states are neglected.}
\label{fig:x2mpl}
\end{figure*}

The emission spectra shown in Fig.~\ref{fig:x2mpl} are calculated for different values of the magnetic field applied along the growth axis and for two possible arrangements of two-electron triplet sublevels. Panel (a) of the Figure~\ref{fig:x2mpl} shows the case of absent structural inversion asymmetry so that the triplet sublevels order is given in~Fig.~\ref{fig:aniso}(a). Panel (b) corresponds to the case of substantial bulk and structural inversion asymmetries where the  triplet sublevels order is given in~Fig.~\ref{fig:aniso}(b). In our calculations the transitions to the singlet two-electron states are disregarded since they are shifted in energy. We also neglect the splittings of initial states due to low symmetry of the QD.

At a zero magnetic field three lines are clearly seen which corresponds to the transitions from $|F_z|=2$ states to the states $|x'\rangle$, $|y'\rangle$ and to the transition from the $|F_z|=1$ states to the state $|0\rangle$ (Fig.~\ref{fig:x2-}). In relatively strong magnetic field four lines are visible. Their order and splittings is determined by the $g$-factors of carriers and by the order of two-electron final states. The emission spectra shown in Fig.~\ref{fig:x2mpl} are similar to those measured in~\cite{kul_exp07}, the detailed comparison of the experimental results and theory will be reported elsewhere.

\section{Conclusions}

To summarize, a theory of spin-orbit induced terms in electron-electron interactions for the carriers confined in $[001]$-grown zinc-blende lattice based quantum wells and quantum dots is developed. Terms originating from the bulk inversion asymmetry and structural inversion asymmetry are derived. 

The theory is applied to calculate the fine structure of two-electron states confined in a single or double QD. In anisotropic systems the spin-orbit electron-electron exchange interaction is shown to lift completely the spin degeneracy of two-electron triplet states. The analytical results are obtained for the parabolic single and lateral double QDs.

We have addressed theoretically the emission spectra of two specific systems where the spin-orbit electron-electron exchange interaction may play an important role: hot trion and $X^{2-}$ complex. In the former case, the two electrons in a triplet state and a hole form an initial state of the complex. In the latter case, the two-electron state is a final one in the process of $X^{2-}$ recombination. The emission spectra of these systems in the magnetic field applied along the growth axis are calculated.

\begin{acknowledgments}
Author thanks V.D. Kulakovskii and E.Ya. Sherman for helpful discussions. The financial support of RFBR, Programmes of RAS and the ``Dynasty'' Foundation -- ICFPM is gratefully acknowledged.
\end{acknowledgments}


\begin{thebibliography}{23}
\expandafter\ifx\csname natexlab\endcsname\relax\def\natexlab#1{#1}\fi
\expandafter\ifx\csname bibnamefont\endcsname\relax
  \def\bibnamefont#1{#1}\fi
\expandafter\ifx\csname bibfnamefont\endcsname\relax
  \def\bibfnamefont#1{#1}\fi
\expandafter\ifx\csname citenamefont\endcsname\relax
  \def\citenamefont#1{#1}\fi
\providecommand{\bibinfo}[2]{#2}
\providecommand{\eprint}[2][]{\url{#2}}

\bibitem[{\citenamefont{Kusraev and Landwehr}(2008)}]{ssc:optor}
\bibinfo{editor}{\bibfnamefont{Y.}~\bibnamefont{Kusraev}} \bibnamefont{and}
  \bibinfo{editor}{\bibfnamefont{G.}~\bibnamefont{Landwehr}}, eds.,
  \emph{\bibinfo{title}{Semicond. Sci. Technol., Special Issue:
  Optical Orientation}}, vol.~\bibinfo{volume}{23} (\bibinfo{publisher}{IOP
  Publishing}, \bibinfo{year}{2008}).

\bibitem[{\citenamefont{Ivchenko}(2005)}]{ivchenko05a}
\bibinfo{author}{\bibfnamefont{E.~L.} \bibnamefont{Ivchenko}},
  \emph{\bibinfo{title}{Optical Spectroscopy of Semiconductor Nanostructures}}
  (\bibinfo{publisher}{Alpha Science, Harrow UK}, \bibinfo{year}{2005}).

\bibitem[{\citenamefont{Bracker et~al.}(2008)\citenamefont{Bracker, Gammon, and
  Korenev}}]{bracker08}
\bibinfo{author}{\bibfnamefont{A.~S.} \bibnamefont{Bracker}},
  \bibinfo{author}{\bibfnamefont{D.}~\bibnamefont{Gammon}}, \bibnamefont{and}
  \bibinfo{author}{\bibfnamefont{V.~L.} \bibnamefont{Korenev}},
  \bibinfo{journal}{Semicond. Sci. Technol.}
  \textbf{\bibinfo{volume}{23}}, \bibinfo{pages}{114004}
  (\bibinfo{year}{2008}).

\bibitem[{\citenamefont{Kusrayev}(2008)}]{kusraev08}
\bibinfo{author}{\bibfnamefont{Y.~G.} \bibnamefont{Kusrayev}},
  \bibinfo{journal}{Semicond. Sci. Technol.}
  \textbf{\bibinfo{volume}{23}}, \bibinfo{pages}{114013}
  (\bibinfo{year}{2008}).

\bibitem[{\citenamefont{Cortez et~al.}(2002)\citenamefont{Cortez, Krebs,
  Laurent, Senes, Marie, Voisin, Ferreira, Bastard, Gerrard, and
  Amand}}]{cortez02}
\bibinfo{author}{\bibfnamefont{S.}~\bibnamefont{Cortez}},
  \bibinfo{author}{\bibfnamefont{O.}~\bibnamefont{Krebs}},
  \bibinfo{author}{\bibfnamefont{S.}~\bibnamefont{Laurent}},
  \bibinfo{author}{\bibfnamefont{M.}~\bibnamefont{Senes}},
  \bibinfo{author}{\bibfnamefont{X.}~\bibnamefont{Marie}},
  \bibinfo{author}{\bibfnamefont{P.}~\bibnamefont{Voisin}},
  \bibinfo{author}{\bibfnamefont{R.}~\bibnamefont{Ferreira}},
  \bibinfo{author}{\bibfnamefont{G.}~\bibnamefont{Bastard}},
  \bibinfo{author}{\bibfnamefont{J.-M.} \bibnamefont{Gerrard}},
  \bibnamefont{and} \bibinfo{author}{\bibfnamefont{T.}~\bibnamefont{Amand}},
  \bibinfo{journal}{Phys. Rev. Lett.} \textbf{\bibinfo{volume}{89}},
  \bibinfo{pages}{207401} (\bibinfo{year}{2002}).

\bibitem[{\citenamefont{Ediger et~al.}(2007)\citenamefont{Ediger, Bester,
  Gerardot, Badolato, Petroff, Karrai, Zunger, and Warburton}}]{ediger:036808}
\bibinfo{author}{\bibfnamefont{M.}~\bibnamefont{Ediger}},
  \bibinfo{author}{\bibfnamefont{G.}~\bibnamefont{Bester}},
  \bibinfo{author}{\bibfnamefont{B.~D.} \bibnamefont{Gerardot}},
  \bibinfo{author}{\bibfnamefont{A.}~\bibnamefont{Badolato}},
  \bibinfo{author}{\bibfnamefont{P.~M.} \bibnamefont{Petroff}},
  \bibinfo{author}{\bibfnamefont{K.}~\bibnamefont{Karrai}},
  \bibinfo{author}{\bibfnamefont{A.}~\bibnamefont{Zunger}}, \bibnamefont{and}
  \bibinfo{author}{\bibfnamefont{R.~J.} \bibnamefont{Warburton}},
  \bibinfo{journal}{Phys. Rev. Lett.} \textbf{\bibinfo{volume}{98}},
  \bibinfo{eid}{036808} (\bibinfo{year}{2007}).

\bibitem[{\citenamefont{Glazov and Kulakovskii}(2009)}]{glazov2009}
\bibinfo{author}{\bibfnamefont{M.~M.} \bibnamefont{Glazov}} \bibnamefont{and}
  \bibinfo{author}{\bibfnamefont{V.~D.} \bibnamefont{Kulakovskii}},
  \bibinfo{journal}{Phys. Rev. B}
  \textbf{\bibinfo{volume}{79}}, \bibinfo{eid}{195305}
  (\bibinfo{year}{2009}).

\bibitem[{\citenamefont{Ivchenko and Pikus}(1997)}]{ivchenkopikus}
\bibinfo{author}{\bibfnamefont{E.~L.} \bibnamefont{Ivchenko}} \bibnamefont{and}
  \bibinfo{author}{\bibfnamefont{G.~E.} \bibnamefont{Pikus}},
  \emph{\bibinfo{title}{Superlattices and other heterostructures}}
  (\bibinfo{publisher}{Springer}, \bibinfo{year}{1997}).

\bibitem[{\citenamefont{Knap et~al.}(1996)\citenamefont{Knap, Skierbiszewski,
  Zduniak, Litwin-Staszewska, Bertho, Kobbi, Robert, Pikus, Pikus, Iordanskii
  et~al.}}]{Knap}
\bibinfo{author}{\bibfnamefont{W.}~\bibnamefont{Knap}},
  \bibinfo{author}{\bibfnamefont{C.}~\bibnamefont{Skierbiszewski}},
  \bibinfo{author}{\bibfnamefont{A.}~\bibnamefont{Zduniak}},
  \bibinfo{author}{\bibfnamefont{E.}~\bibnamefont{Litwin-Staszewska}},
  \bibinfo{author}{\bibfnamefont{D.}~\bibnamefont{Bertho}},
  \bibinfo{author}{\bibfnamefont{F.}~\bibnamefont{Kobbi}},
  \bibinfo{author}{\bibfnamefont{J.~L.} \bibnamefont{Robert}},
  \bibinfo{author}{\bibfnamefont{G.~E.} \bibnamefont{Pikus}},
  \bibinfo{author}{\bibfnamefont{F.~G.} \bibnamefont{Pikus}},
  \bibinfo{author}{\bibfnamefont{S.~V.} \bibnamefont{Iordanskii}},
  \bibnamefont{V. Mosser, K. Zekentes, and Yu. B. Lyanda-Geller}, \bibinfo{journal}{Phys. Rev. B}
  \textbf{\bibinfo{volume}{53}}, \bibinfo{pages}{3912} (\bibinfo{year}{1996}).

\bibitem{lawaetz71} P. Lawaetz, Phys. Rev. B {\bf 4}, 3460 (1971).

\bibitem[{\citenamefont{Berestetskii et~al.}(1999)\citenamefont{Berestetskii,
  Pitaevskii, and Lifshitz}}]{ll4_eng}
\bibinfo{author}{\bibfnamefont{V.~B.} \bibnamefont{Berestetskii}},
  \bibinfo{author}{\bibfnamefont{L.~P.} \bibnamefont{Pitaevskii}},
  \bibnamefont{and} \bibinfo{author}{\bibfnamefont{E.~M.}
  \bibnamefont{Lifshitz}}, \emph{\bibinfo{title}{Quantum Electrodynamics,
  Second Edition (vol. 4)}} (\bibinfo{publisher}{Butterworth-Heinemann,
  Oxford}, \bibinfo{year}{1999}).

\bibitem[{\citenamefont{Glazov and Ivchenko}(2004)}]{glazov04a}
\bibinfo{author}{\bibfnamefont{M.~M.} \bibnamefont{Glazov}} \bibnamefont{and}
  \bibinfo{author}{\bibfnamefont{E.~L.} \bibnamefont{Ivchenko}},
  \bibinfo{journal}{JETP} \textbf{\bibinfo{volume}{99}}, \bibinfo{pages}{1279}
  (\bibinfo{year}{2004}).

\bibitem[{\citenamefont{Averkiev et~al.}(2002)\citenamefont{Averkiev, Golub,
  and Willander}}]{averkiev02}
\bibinfo{author}{\bibfnamefont{N.~S.} \bibnamefont{Averkiev}},
  \bibinfo{author}{\bibfnamefont{L.~E.} \bibnamefont{Golub}}, \bibnamefont{and}
  \bibinfo{author}{\bibfnamefont{M.}~\bibnamefont{Willander}},
  \bibinfo{journal}{J. Phys.: Condens. Matter} \textbf{\bibinfo{volume}{14}},
  \bibinfo{pages}{R271} (\bibinfo{year}{2002}).

\bibitem[{\citenamefont{Tarasenko and Ivchenko}(2005)}]{tarasenko05eng}
\bibinfo{author}{\bibfnamefont{S.~A.} \bibnamefont{Tarasenko}}
  \bibnamefont{and} \bibinfo{author}{\bibfnamefont{E.~L.}
  \bibnamefont{Ivchenko}}, \bibinfo{journal}{JETP Letters}
  \textbf{\bibinfo{volume}{81}}, \bibinfo{pages}{231} (\bibinfo{year}{2005}).

\bibitem[{\citenamefont{Badalyan and Vignale}(2009)}]{Badalyan-2009}
\bibinfo{author}{\bibfnamefont{S.~M.} \bibnamefont{Badalyan}} \bibnamefont{and}
  \bibinfo{author}{\bibfnamefont{G.}~\bibnamefont{Vignale}},
  \bibinfo{journal}{arXiv:0907.2995}  (\bibinfo{year}{2009}).

\bibitem[{\citenamefont{Bir and Ivchenko}(1975)}]{birivch75}
\bibinfo{author}{\bibfnamefont{G.~L.} \bibnamefont{Bir}} \bibnamefont{and}
  \bibinfo{author}{\bibfnamefont{E.~L.} \bibnamefont{Ivchenko}},
  \bibinfo{journal}{Sov. Phys. Semicond.} \textbf{\bibinfo{volume}{9}},
  \bibinfo{pages}{858} (\bibinfo{year}{1975}).

\bibitem[{\citenamefont{Aleiner and Ivchenko}(1992)}]{aleiner92eng}
\bibinfo{author}{\bibfnamefont{I.~L.} \bibnamefont{Aleiner}} \bibnamefont{and}
  \bibinfo{author}{\bibfnamefont{E.~L.} \bibnamefont{Ivchenko}},
  \bibinfo{journal}{JETP Letters} \textbf{\bibinfo{volume}{55}},
  \bibinfo{pages}{692} (\bibinfo{year}{1992}).

\bibitem[{\citenamefont{Goupalov et~al.}(1998)\citenamefont{Goupalov, Ivchenko,
  and Kavokin}}]{goupalov98}
\bibinfo{author}{\bibfnamefont{S.~V.} \bibnamefont{Goupalov}},
  \bibinfo{author}{\bibfnamefont{E.~L.} \bibnamefont{Ivchenko}},
  \bibnamefont{and} \bibinfo{author}{\bibfnamefont{A.~V.}
  \bibnamefont{Kavokin}}, \bibinfo{journal}{JETP}
  \textbf{\bibinfo{volume}{86}}, \bibinfo{pages}{388} (\bibinfo{year}{1998}).

\bibitem[{\citenamefont{Glazov et~al.}(2007)\citenamefont{Glazov, Ivchenko, von
  Baltz, and Tsitsishvili}}]{glazov2007a}
\bibinfo{author}{\bibfnamefont{M.~M.} \bibnamefont{Glazov}},
  \bibinfo{author}{\bibfnamefont{E.~L.} \bibnamefont{Ivchenko}},
  \bibinfo{author}{\bibfnamefont{R.}~\bibnamefont{von Baltz}},
  \bibnamefont{and} \bibinfo{author}{\bibfnamefont{E.~G.}
  \bibnamefont{Tsitsishvili}}, \bibinfo{journal}{Int. J. Nanoscience}
  \textbf{\bibinfo{volume}{6}}, \bibinfo{pages}{265} (\bibinfo{year}{2007}).

\bibitem[{\citenamefont{\c{S}. C.~B\u{a}descu et~al.}(2005)\citenamefont{\c{S}.
  C.~B\u{a}descu, Lyanda-Geller, and Reinecke}}]{badescu:161304}
\bibinfo{author}{\bibnamefont{\c{S}. C.~B\u{a}descu}},
  \bibinfo{author}{\bibfnamefont{Y.~B.} \bibnamefont{Lyanda-Geller}},
  \bibnamefont{and} \bibinfo{author}{\bibfnamefont{T.~L.}
  \bibnamefont{Reinecke}}, \bibinfo{journal}{Phys. Rev. B} \textbf{\bibinfo{volume}{72}},
  \bibinfo{eid}{161304} (\bibinfo{year}{2005}).
  
\bibitem[{\citenamefont{Kavokin}(2001)}]{kkavokin01}
\bibinfo{author}{\bibfnamefont{K.~V.} \bibnamefont{Kavokin}},
  \bibinfo{journal}{Phys. Rev. B} \textbf{\bibinfo{volume}{64}},
  \bibinfo{pages}{075305} (\bibinfo{year}{2001}); 
\bibinfo{author}{\bibfnamefont{K.~V.} \bibnamefont{Kavokin}},
  \bibinfo{journal}{Phys. Rev. B} \textbf{\bibinfo{volume}{69}},
  \bibinfo{pages}{075302} (\bibinfo{year}{2004}).

\bibitem[{\citenamefont{Gangadharaiah et~al.}(2008)\citenamefont{Gangadharaiah,
  Sun, and Starykh}}]{gangadharaiah:156402}
\bibinfo{author}{\bibfnamefont{S.}~\bibnamefont{Gangadharaiah}},
  \bibinfo{author}{\bibfnamefont{J.}~\bibnamefont{Sun}}, \bibnamefont{and}
  \bibinfo{author}{\bibfnamefont{O.~A.} \bibnamefont{Starykh}},
  \bibinfo{journal}{Phys. Rev. Lett.} \textbf{\bibinfo{volume}{100}},
  \bibinfo{eid}{156402} (\bibinfo{year}{2008}).

\bibitem[{\citenamefont{Chekhovich
  et~al.}(2007{\natexlab{a}})\citenamefont{Chekhovich, Brichkin, Chernenko,
  Kulakovskii, Sedova, Sorokin, and Ivanov}}]{kul1:07}
\bibinfo{author}{\bibfnamefont{E.~A.} \bibnamefont{Chekhovich}},
  \bibinfo{author}{\bibfnamefont{A.~S.} \bibnamefont{Brichkin}},
  \bibinfo{author}{\bibfnamefont{A.~V.} \bibnamefont{Chernenko}},
  \bibinfo{author}{\bibfnamefont{V.~D.} \bibnamefont{Kulakovskii}},
  \bibinfo{author}{\bibfnamefont{I.~V.} \bibnamefont{Sedova}},
  \bibinfo{author}{\bibfnamefont{S.~V.} \bibnamefont{Sorokin}},
  \bibnamefont{and} \bibinfo{author}{\bibfnamefont{S.~V.}
  \bibnamefont{Ivanov}}, \bibinfo{journal}{Phys. Rev. B}
  \textbf{\bibinfo{volume}{76}}, \bibinfo{pages}{165305}
  (\bibinfo{year}{2007}{\natexlab{a}}).

\bibitem[{\citenamefont{Chekhovich
  et~al.}(2007{\natexlab{b}})\citenamefont{Chekhovich, Brichkin, Chernenko, and
  Kulakovskii}}]{kul_exp07}
\bibinfo{author}{\bibfnamefont{E.~A.} \bibnamefont{Chekhovich}},
  \bibinfo{author}{\bibfnamefont{A.~S.} \bibnamefont{Brichkin}},
  \bibinfo{author}{\bibfnamefont{A.~V.} \bibnamefont{Chernenko}},
  \bibnamefont{and} \bibinfo{author}{\bibfnamefont{V.~D.}
  \bibnamefont{Kulakovskii}}, in \emph{\bibinfo{booktitle}{Proc. 15th Int.
  Symp. "Nanostructures: Physics and Technology", Novosibirsk, Russia}}
  (\bibinfo{year}{2007}{\natexlab{b}}).

\end{thebibliography}
\end{document}